\documentclass[a4paper]{jpconf}

\usepackage{graphicx}

\newcommand{\Eq}[1]{Eq. (\ref{#1})}

\begin{document}
\title{Symmetry breaking as the origin of zero-differential
resistance states of a  2DEG  in strong  magnetic fields.}

\author{Manuel Torres $\null^{1}$ and Alejandro Kunold   $\null^{2}$  }

\address{  $\null^{1}$  Instituto de F\'{\i}sica,
  Universidad Nacional Aut\'onoma de M\'exico, 01000, M\'exico \\
 $\null^{2}$  Departamento de Ciencias B\'asicas,
Universidad Aut\'onoma Metropolitana-Azcapotzalco,
 02200, M\'exico}

\ead{torres@fisica.unam.mx}

\begin{abstract}

Zero resistance differential states have been  observed in two-dimensional
electron gases  (2DEG)  subject  to  a  magnetic field and a  strong   dc current. 
In a recent work we presented   a model to describe  the nonlinear transport regime of this phenomenon. From  the analysis of the differential resistivity and the longitudinal voltage we predicted the formation of negative differential resistivity states, although these states are known to be  unstable.
Based on our  model,   we derive  an  analytical approximated  expression for the Voltage-Current characteristics, that captures the main elements of the problem. 
The result allow us to  construct  an energy  functional  for  the system. In the zero temperature limit, the system 
presents a quantum phase transition,  with the control parameter given by the magnetic field. 
It is noted that above  a threshold value    ($B   > B_{th}$),  the symmetry is spontaneously broken. 
At sufficiently high magnetic field and low temperature  the  model predicts a phase with
a  non-vanishing permanent current; this is 
a novel phase that  has not been observed so far.  
 
\end{abstract}

\section{Introduction}

The study of non-equilibrium magnetotransport in high mobility two-dimensional
electron gases  (2DEG)  has acquired great experimental and theoretical interest.
Strong magnetoresistance  oscillations (MIRO)  and zero resistance states (ZRS)  have been observed  in 
high mobility  \\ $GaAs/Al_xGa_{1-x} As$  heterostructures subject  to a magnetic field and to microwave radiation \cite{zudov:201311,mani:646}. 
Our current understanding
of this phenomenon rests upon models that predict the existence
of negative-resistance states (NRS) yielding an instability that
rapidly drives the system into a ZRS \cite{andreev:056803}. Two distinct mechanisms
for the generation of NRS are known, one is based in the
microwave-induced impurity scattering
\cite{ryzhii:2078,durst:086803,lei:226805,vavilov:035303,
torres:115313, inarrea:073311}, while the second
is linked to inelastic processes leading to  a non-trivial  electron distribution function 
 \cite{dmitriev:226802,dmitriev:115316,robinson:036804}.

More recently, Hall field-induced resistance oscillations (HIRO)   and zero differential resistance states (ZDRS)  have 
 been observed in response to strong  dc electric current  excitations   \cite{bykov:245307, zhang:081305,zhang:041304}. The effect of a direct dc  current 
on electron transport  can be quite dramatic leading to zero 
differential resistance states (ZDRS)\cite{bykov:116801}.
 At low temperature and above a threshold bias current
the differential resistance vanishes and the
longitudinal dc voltage becomes constant.    Bykov et al.\cite{bykov:116801} discussed  their experimental results following  an approach 
similar to that
of Andreev et al. \cite{andreev:056803}.  In terms of the differential longitudinal resistivity 
$\rho_{xx}  $ the stability condition reads
\begin{equation}\label{stabcond}
\rho_{xx} =  \frac{\partial E_x}{\partial J_x}   \ge 0 \, .
\end{equation}
Hence, according to the condition in Eq. (\ref{stabcond}) the 2DES is unstable  at negative 
differential resistance. The presence of the ZDRS can be  attributed to
 the formation of negative differential resistance states (NDRS) that yields an instability that drives
the system into a ZDRS. 

\section{Model and results}

In a recent work we presented   a model \cite{kunold:3803,kunold:205314} to describe the nonlinear response to a direct  dc current
applied to a 2DEG in a strong  magnetic field.
The model incorporates the exact dynamics of two-dimensional damped electrons in the
presence of arbitrarily strong  magnetic  and  dc electric fields,
while the effects of randomly distributed impurities are perturbatively added.
From the analysis of the differential resistivity and the longitudinal voltage
we observe the formation of negative differential resistivity states.
Both the effects of elastic impurity scattering
as well as those related to inelastic processes play an important role.
The theoretical predictions  correctly reproduce the
main  experimental features  provided that the inelastic scattering  rate  obeys a
$T^2$ temperature dependence, consistent with electron-electron interaction effects.

The more relevant results are presented here.
After  the  current density is worked out, it splits into  a Drude and an impurity 
 induced contribution: $ \mathbf   {J}= \mathbf   {J}^D + \mathbf   {J}^{imp}$.
The Drude contribution is given  by  $\mathbf   {J}^D  = n_e \, e \, \mathbf   {v}_d $, where $n_e$ is the electron density and    the drift velocity  is given as

\begin{equation}\label{vderiva}
 \mathbf   {v}_d =  \frac{ e\tau}{m^*} \, \frac{\mathbf   {E }-\omega_c\tau \, \mathbf   {e}_z \times \mathbf   {E}}
{1+\omega_c^2\tau_{tr}^2} \, .
\end{equation}  
Here, $e_z$ is the unit vector normal to the plane of the system, $\omega_c$ is the cyclotron frequency,   $m^*$ is the effective mass of the charge carrier, and   $\tau_{tr}$ is the transport scattering time.  It is important to point out that  
the Drude term  is not added by hand; instead it explicitly appears in our formalism  because the exact  solution of the 
Schr\"odinger equation in the presence of   magnetic an electric fields is  obtained
 from a unitary transformation \cite{kunold:205314}  that incorporates  the
 solution of the classical equation of motion. 
   
  In the presence of elastic  scattering an electron exchanges a momentum $  \mathbf   {q} =  \mathbf   {k}  -  \mathbf   {k}^\prime$
  with the impurity scatterers. Hence,  the  components of  the  impurity induced density current can be expressed  as
\begin{equation}\label{dencu2}
 J^{imp}_i= 
\frac{e \, \omega_c\, n_{imp} }{\hbar^2}\sum_{\mu\mu^{\prime}}
\int \frac{d^2q}{2 \pi}
\left[f\left( {\cal E}_{\mu,\frac{q}{2}}\right)-f\left( {\cal E}_{\mu^\prime,-\frac{q}{2}}\right)\right]
G^i_{\mu\mu^{\prime}}
\left(q\right) \, ,
\end{equation}
where $i=x,y$;  $n_{imp} $ is the impurity density, and $f$  is the Fermi  distribution  function evaluated at the energies
 ${\cal E}_{\mu, q/2}$ and  $ {\cal E}_{\mu^\prime, -q/2}$  respectively.  
 The   Landau levels are  tilted by the electric field according to   $ {\cal E}_{\mu,  k} = \hbar \omega_c 
 \left( \mu + \frac{1}{2} \right) +  \hbar \omega_k$; with  $ \omega_k = \mathbf   {k}  \cdot \mathbf   {v}_d$. 
 Notice that the $x$ component of the exchanged momentum   $  q_x =  k_x  -  k_x^\prime$
 is equivalent to a hopping (shifting of the guiding center) in the $y$ direction $\Delta y = l_B^2 q_x$; with 
 $l_B = \sqrt{\hbar / e B}$ the magnetic length. The function $G^i_{\mu\mu^{\prime}}$ in \Eq{dencu2}  is given by 
\begin{equation}\label{dosexy}
G^i_{\mu\mu^{\prime}} = \left\vert V\left(q\right) \right\vert^2
\left\vert D_{\mu\mu^{\prime}}
\left(z_q\right)\right\vert^2 
\times\frac{q_i \Delta_{\mu\mu^{\prime}}
+2\left\vert\epsilon_{ij}\right\vert q_j\omega_c \Gamma}
{\Delta^2_{\mu\mu^{\prime}}+4\omega_c^2\Gamma^2},
\end{equation}
in the previous equation   $\epsilon_{i,j}$ is the 2D antisymmetric tensor,  $\Delta_{\mu\mu^{\prime}}
 =\left[\omega_q+\omega_c\left(\mu-\mu^{\prime}\right)\right]^2
-\omega_c^2+\Gamma^2 $,
 $V(q) $ is the Fourier component of the scattering potential, and $\Gamma$ is the Landau level broadening factor.
 In order to take into  account the
known fact that the width of LLs depends on the magnetic field \cite{ando:437},
 we  consider $ \Gamma^2 = \beta  (2 \omega_c / \pi \tau_{tr} )$ where $\beta $ is a
phenomenological parameter that takes into account the  difference between the transport scattering time $\tau_{tr}$ and the quantum scattering time $\tau_q$.  In the case of  short range 
neutral impurities $\beta = 1$. 
 The  matrix elements $D_{\mu,\nu}$ are given in terms of the  associated Laguerre polynomial, see \cite{kunold:205314}.

In recent work is has been reported that the temperature dependence of nonlinear oscillatory magnetoresistance in 2DEG subject to a strong dc  electric field can be explained  if the 
quantum scattering time $\tau_q$ incorporates  a temperature dependence that is attributed to electron-electron interactions  \cite{hatke:161308}. The quantum scattering rate is written as 

\begin{equation}\label{trelaj}
\frac{1}{ \tau_{q}}  = \frac{1}{ \tau_{imp} } + \frac{1}{ \tau_{ee}} \,,  \hskip 1.5cm 
\frac{1}{  \tau_{ee} } = \frac{ \lambda 
  \left( k_B T\right)^2 }{ \hbar  \,  E  _F} \, , 
\end{equation}
 the parameter $\lambda $ has to be  experimentally determined, but it  is a constant of the order 
 of the unity.  For short range 
neutral scatterers  the impurity  rate can be estimated as
$1/\tau_{imp}=4\pi^2\hbar\alpha n_{imp}/m^*$,  where $\alpha$ is  related to the scattering length \cite{torres:115313}.

Let us now consider the nonlinear transport regime.  In a current controlled
scheme: the longitudinal density current is fixed to
a constant value $J_{dc}$ and there is no transverse current, $J_y$.
This leads to a set of two implicit equations for
the density current
 \begin{eqnarray}\label{curr2}
      J_x\left(E_x,E_y\right) & =& J_{dc}   \,  ,     \nonumber \\
     J_y\left(E_x,Ey\right) & =& 0  \, . 
\end{eqnarray}
 They represent two implicit equation for the unknown   $E_x$ and  $E_y$;
 the equations can  be solved following a self-consistent iteration \cite{kunold:205314}.
 However, it is  verified  that for the conditions that apply  in  experiments and in the separated LL regions ($\omega_c \tau_{tr} \gg 1$),  the solution of the previous equations simplify   because the 
 following conditions  $E_x \ll E_y$ and  $ J_x^{imp}     \ll    e n_e E_y /B $
 are  simultaneously satisfied.  Hence,  from the first equation in (\ref{curr2})  it follows
 that the leading contribution to the Hall electric field is given by the classical 
 result  $   E_y =  B J_{dc}  / e n_e  $.
 Whereas the second equation in (\ref{curr2}) yields  a nonlinear $E_x \, - \, J_{dc}$ relation

 \begin{equation}\label{relej}
E_x = \frac{m^*}{e^2 \, n_e \tau_{tr} } \, J_{dc}  + \frac{B}{e n_e} \, J^{imp}_y \left( E_x \sim 0, E_y \sim  \frac{B J_{dc}  }{e n_e} \right)   \, .
\end{equation}
 The expression for the impurity induced currents $J^{imp}_y $  is given in \Eq{dencu2}. In what follows we set $ J_{dc} \equiv J_x$.
 
We had previously  presented numerical results  in very good agreement with the experimental ones, based in \Eq{relej} 
\cite{kunold:3803}, and 
in the self consistent solution of equations in (\ref{curr2}) \cite{kunold:205314}.
 However in this work we point out that an approximated analytical  $E_x- J_x$  relation  can be obtained. 
As  we are interested in the separated LL region  the dominant effect is provided by the intra-Landau transitions.   
  We include the following assumptions: (1)   only the  LL  nearest to the Fermi level  contribute. (2) The potential is taken as a constant $V(q) \equiv 
  2 \pi \hbar^2  \alpha / m^* $,  corresponding to a short-range impurity scattering,  and (3) the integrand in \Eq{dencu2} has a  dominant 
  peak at the value $q_{max} \approx \sqrt{ 8 \pi \, n_e}$, this allow us to carry out explicitly the integral over the Laguerre polynomials. After the angular integral are evaluated, a simply    $E_x \, - \, J_x$ relation is obtained

   \begin{equation}\label{relej1}
 E_x \, = \,   C_1  \, \frac{J_x }{\tau_{tr}}   + \,  C_2 \,  \frac{B^{3/2}} {\tau_{tr} \, J_x} \, 
 \left[ 1 - \frac{1}{\sqrt{1 +  \frac{ C_3 \, \tau_{tr} \,   J_x^2}{B}}}\right]   \, ,
\end{equation}
the   parameters in this equation are given as  
 \begin{equation}\label{param1}
C_1 =\frac{m^*}{e^2 n_e}, \hskip0.5cm
 \hskip0.9cm
C_2 =  \hbar  n_{imp}  \,  \alpha^2  \left( \frac{2 \pi  e}{m^*} \right)^{3/2} ,
\hskip0.9cm
C_3 =  \frac{2 \pi^2  m^*}{e^3 n_e  } \, .
\end{equation}

\Eq{relej1} depends on the   a temperature  through the $T$-dependence of $\tau_{tr}$. In the simplest approximation, short range impurity scatterers, we can consider that  $\tau_{tr}$ is proportional to $\tau_q$, 
consequently  the $T$ dependence is dictated by \Eq{trelaj}. 
Utilizing  this assumption and the   simple equation in  (\ref{trelaj}) we carried out   the comparison with the experimental results, observing a very good agreement. In particular the transition from 
positive to zero differential resistance states shows the correct dependence on the temperature 
\cite{kuntor:001}.

In this work  we shall  concentrate in a simple but interesting limit, we take $T=0$ and consider   the analysis of the stability of the NDRS. Therefore  the transport scattering rate coincides with 
the impurity scattering rate: $1/\tau_{tr} = 1/\tau_{imp}=4\pi^2\hbar\alpha n_{imp}/m^*$.
In the region of positive differential resistivity $\rho_{xx} = \frac{\partial E_x}{\partial J_x} > 0$, the relation between between $E_x$ 
and $J_x$ is single-valued  and an homogeneous uniform electron density   throughout the system is stable. 
On the other hand,  in the   region  in which   $\rho_{xx}$ is negative corresponding to a NDRS,  the system becomes unstable. An approximated scheme (Maxwell construction) replaces the negative slope
by  a horizontal line, $i.e.$ a ZDRS. The real reason for the appearance  of a negative slope in the
 $E_x - J_x$  plot  is the implicit restraint of uniform electron density throughout the system.
 However  in these regions configurations with a non uniform current distribution 
  turn out to be the equilibrium configuration of the system.  The simplest possible pattern is a domain wall: two parts of the sample carry stable density currents $J_{x1}$ and 
  $J_{x2}$ with the same value of $E_x$. Hence we expect a phase transition defined  by a critical iso-$B$ line, that passes through a  
   critical point  that is a point of inflection of this iso-$B$ line; so both  
  $ \frac{\partial E_x}{\partial J_x}  $ and  $ \frac{\partial^2 E_x}{\partial J_x^2}  $ vanish at this point. The critical  values: $E_x = E_c$,  $J_x = J_c$  and $B=B_c$; at which these conditions are satisfied are readily determined as 
    
   \begin{equation}\label{param2}
   J_c =  \xi_1 \frac{C_1}{\tau_{imp}^2 \,  C_2 \,  C_3^{3/2}}, \hskip0.8cm    E_c =  \xi_2 \frac{C_1^2}{\tau_{imp}^3 \,  C_2 \,  C_3^{3/2}}
\,  , \hskip0.8cm  B_c^{1/2} =   \xi_3  \frac{C_1}{\tau_{imp}^{3/2} \,  C_2 \,  C_3}\, .
\end{equation}   
  with  the values of the constants $\xi_i$ given by: $ \xi_1 \approx 40.7 \, , \hskip0.2cm      \xi_2 \approx 161.9  \, ,  \hskip0.2cm  and  \hskip0.2cm
\,   \xi_3 \approx 20.85 \, .$  Notice that the method follows similar steps those  used in the analysis of the critical point of the Van der  Waals equation \cite{pathria:1996}. 
  
 Using Eqs.  (\ref{param1}),  and  (\ref{param2})    we can express $J_c$, $E_c$ and $B_c$ is terms of the parameters 
of the system. In order to  get an  estimation  we consider:  $n_e  = 8 \times 10^{15}m^{-2}$,   $\mu=100 m^2 V/s$, 
$\alpha = 0.1$, $n_{imp}  = 4.5 \times 10^{11}m^{-2}$.  So we obtain:
 
\begin{equation}\label{paramcrit}
   J_c \sim  0.4   \, A/m      \,  , \hskip1.0cm    E_c  \sim  6.5  \,  V/m \, ,   \hskip1.0cm  B_c \sim  \, 0.12  \, T \,   .
\end{equation}
These values correspond to zero temperature, so  are not directly comparable with the experimental ones, however 
the order of magnitude show  a reasonable agreement \cite{bykov:116801}. 

We now define reduced variables 

 \begin{equation}\label{paramred}
   J_r = \frac{J_x}{J_C} \, ,  \hskip1.0cm    E_r = \frac{E_x}{E_C} 
\,  , \hskip1.0cm  B_r = \frac{B}{B_C} \, .
\end{equation}  
Hence the critical point is localized at $J_r = E_r = B_r = 1$. Using (\ref{relej1}) and (\ref{paramred}), we readily obtain the reduced voltage-current characteristic

 \begin{equation}\label{relejred}
 E_r \, = \, 0.25  \, J_r   \, + \, \frac{1.38 }{J_r} \, B_r^{3/2}  \, \left( 1 -  \left[ 1 +  3.8 \,  \frac{J_r^2 }{B_r } \right]^{-1/2} \right)  \, ,
\end{equation}
which is   ``universal'', in the sense that  it
does not explicitly contains any parameter of  
the system.

\begin{figure}
\begin{center}
\includegraphics[width=12 cm]{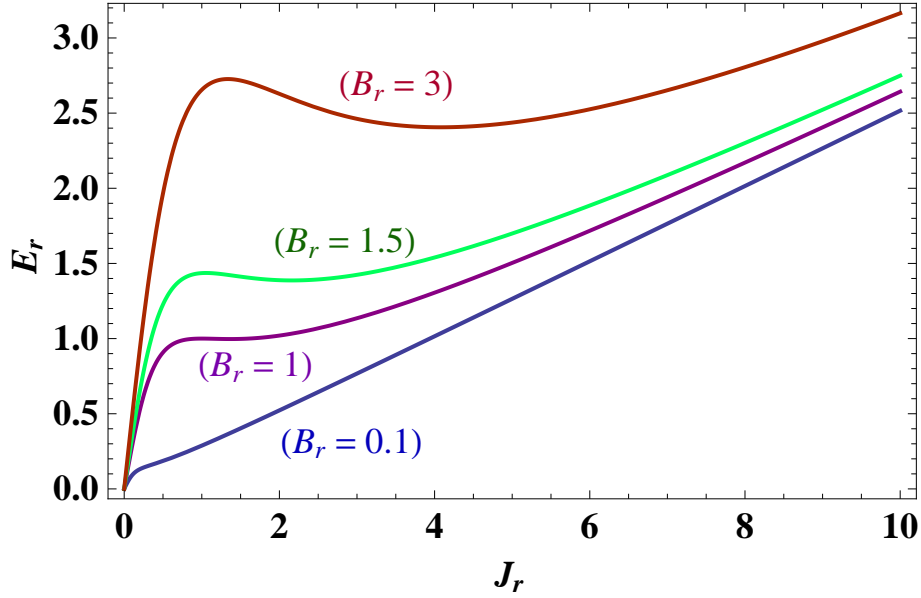}
\end{center}
\caption{  (Color on line). Reduced longitudinal electric field $E_r$ as function of the reduced longitudinal current $J_r$, the values 
of $B_r $ are displayed  near to each line.}
\end{figure}

 Fig. 1  display a series of  plots of  reduced electric field  $E_r$  as a function of the reduced 
longitudinal density current $J_r$ ; for various  values of the reduced magnetic field. We clearly observe a phase  transition:
below the threshold $ B_r < 1$ the slope of the curve is always positive and the system is  stable.
On the other hand if $ B_r  > 1$  and above the  threshold bias current $(J_r > 1)$
the differential resistivity becomes negative.

We can define a scalar Lyapunov functional  $W [ J_r] = \int  \,  d^2 r \,  w \left[  J_r \right] $, 
where 
 \begin{equation}\label{LF1}
 w [  J_r]  = \int^{E_r}  J_r \left[E_r^\prime \right] \,  d E_r^\prime   \,    =  \,   \int^{J_r}  J \,^\prime  \,   \frac{d E_r }{d J_r^\prime  } \, d J_r^\prime   \, . 
\end{equation}

Using \Eq{relejred} a simple calculations yields: 

 \begin{equation}\label{Lyapu}
 w [  J_r]  = 0.125   \,  J_r^2 +  1.38 \, B_r^{3/2} \left[ 1 -  \log \left\{ 2   \left( 1 +  \sqrt{1 +  3.8 \,  \frac{ J_r^2 }{B_r}}         \right)  \right\}  -      \left[ 1 +  3.8 \,  \frac{J_r^2 }{B_r } \right]^{-1/2}
\right]    \, .
\end{equation}

One can show  that $ W[J_r]$ is a Lyapunov functional, $i.e.$ a non-increasing function of time, so  its minima are stable steady states.
To the  expression for $ w [  J_r]$ in \Eq{Lyapu}  one should add a field (current) derivative contribution  of the form $\lambda  \vert \nabla 
\cdot \mathbf{E} \vert^2$, where $\lambda$  introduces a domain wall thickness scale that produces a positive contribution  to $W$.
 Fig. 2  display a series of  plots of   $w$  as a function of $J_r$ 
  for various  values of the reduced magnetic field. If $ B_r < 1$ the functional $W [ J_r] $  has a single minima at $J_r = 0$,
  however when  $ B_r  > 1$  new minima appear at $J_r \ne 0$. Furthermore we observe that for   sufficiently strong magnetic field  the minima at  $J_r = 0$  is not longer  
  the ground state. In general the system  is expected to relax to one of the local minimum of $W,$ not necessarily the ground state. However in the presence of noise the system is expected to escape from the high-lying minima. 
  This suggest that  at sufficiently high magnetic field and low temperature  the model predicts    a phase  with a permanent 
  non-vanishing current $J_m  $, where $J_m$ is the minima of $W$. 
  This   is a novel phase  that has not been observed so far, hence it deserves further studies. 
  In future work we will use the present results to analyze 
  in detail the  formation and structure of the   domain walls and also to include the finite temperature effects.

\begin{figure}
\begin{center}
\includegraphics[width=12 cm]{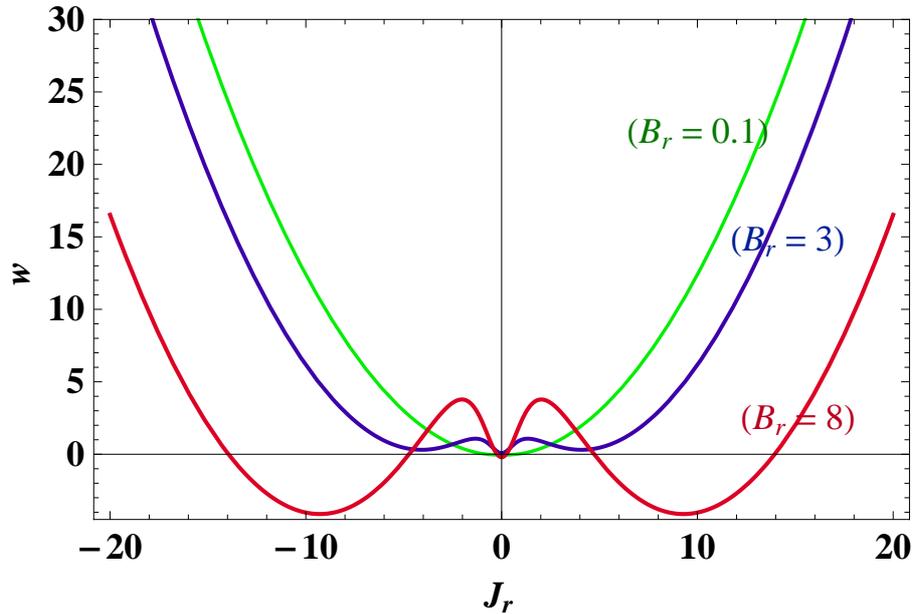}
\caption{ (Color on line).  A plot of $w$ $vs$ $J_r$, for values of $B_r = 0.1,\, 3, \, and \,  8$.}
\end{center}
\end{figure}

  \subsection{Acknowledgments}
We acknowledge support 
from UNAM project  PAPIIT-IN118610.

\end{document}